\documentclass[pra,aps,10pt,showpacs,superscriptaddress,twocolumn]{revtex4-1}
\usepackage{color} 
\usepackage{amssymb} 
\usepackage{amsmath} 
\usepackage{graphicx}
\usepackage{epstopdf}
\usepackage{mathrsfs}
\usepackage{textcomp}
\usepackage{makeidx}
\usepackage{amsthm}
\usepackage{amsfonts}
\usepackage{amssymb}  
\usepackage{hyperref}
\usepackage{times}

\newcommand{\ket}[1]{\vert#1\rangle}

\newcommand{\braket}[2]{\langle#1\vert#2\rangle}
\newcommand{\ketbra}[2]{\vert#1\rangle\langle#2\vert}
\newcommand{\braketM}[3]{\langle#1\vert#2\vert#3\rangle}
\newcommand{\mean}[2]{\langle#1\rangle_{#2}}

\newcommand{\modulo}[1]{\vert#1\vert}

\newcommand{\funcx}[4]{#1_{#3}^{#4}({\bf #2})}
\newcommand{\opx}[4]{\hat{#1}_{#3}^{#4}({\bf #2})}
\newcommand{\opt}[4]{\hat{#1}_{#3}^{#4}(#2)}
\newcommand{\op}[3]{\hat{#1}_{#2}^{#3}}

\begin{document}

\title{Probing mechanical quantum coherence with an ultracold-atom meter}

\author{N. Lo Gullo} 

\author{Th. Busch} \affiliation{Department of Physics, University
  College Cork, Cork, Republic of Ireland}

\author{G. M. Palma} \affiliation{NEST Istituto Nanoscienze-CNR and
  Dipartimento di Fisica, Univerisita' degli Studi di Palermo, Via
  Archirafi 36, I-90123 Palermo, Italy} \date{\today}

\author{M. Paternostro} \affiliation{Centre for Theoretical Atomic,
  Molecular and Optical Physics, School of Mathematics and Physics,
  Queen's University, Belfast BT7 1NN, United Kingdom} \date{\today}

\begin{abstract}
  We propose a scheme to probe quantum coherence in the state of a
  nano-cantilever based on its magnetic coupling (mediated by a
  magnetic tip) with a spinor Bose Einstein condensate (BEC). By
  mapping the BEC into a rotor, its coupling with the cantilever
  results in a gyroscopic motion whose properties depend on the state
  of the cantilever: the dynamics of one of the components of the
  rotor angular momentum turns out to be strictly related to the
  presence of quantum coherence in the state of the cantilever.  We
  also suggest a detection scheme relying on Faraday rotation, which
  produces only a very small back-action on the BEC and it is thus
  suitable for a continuous detection of the cantilever's dynamics.
\end{abstract}

\maketitle


Recently, a considerable research effort has been put in achieving
quantum control of micro and nano-scale mechanical
systems~\cite{espe}.  The role played by such objects in the current
quest for demonstrating quantumness at the mesoscopic scale has
changed in time, and by today they have become key players.  Micro and
nano-mechanical devices are now considered for quantum technological
purposes as well as foundational questions~\cite{espe,cool,cool2}.



As interesting and promising as they could be, such systems are in
general very difficult to probe and measure directly. The necessity of
isolating their fragile dynamics from the influences of the outside
world and the need for low operating temperatures that allow for the
magnification of the quantum mechanical features of their motion often
imply that no direct access to such devices is possible. By today
several schemes exist that use the interaction with light to extract
information from the mechanical structures~\cite{mauro}. However, such
methods are certainly not exhaustive and a more systematic approach to
measure the quantum features of micro/nano-mechanical devices is
highly desirable.


In this sense, a considerable step forward has been the design of
interfaces between mechanical systems and ancillae such as
superconducting systems and (ultra-)cold atomic
ensembles~\cite{tip,mauro2}, which can be used to efficiently monitor,
measure, and prepare the inaccessible mechanical counterparts.  Most
interestingly, some of these hybridization strategies are already
mature enough to have found interesting preliminary
implementations~\cite{hunger}. In this work we present a new strategy
by demonstrating that the interaction between an ultra-cold atomic
system and a mechanical oscillator can be exploited for effective
diagnostics of mechanical quantum coherences. 
A similar approach has been used in a recent work \cite{kind} for 
different purposes. 
Along the lines of
Ref.~\cite{tip}, where it was shown that a similar system can mimic
the strong coupling regime of cavity quantum electrodynamics, we
consider a setup composed of a mechanical oscillator placed on an atom
chip and coupled to a spinor BEC through a magnetic tip. In our
scheme, the magnetic tip acts as a transducer turning the mechanical
oscillations into a magnetic field experienced by the atomic
spins. The motion of the latter in turn results in a driving force for
the mechanical oscillator. A physically transparent description of the
mechanism underlying our proposal is provided by the formal mapping of
the spinor BEC onto a tridimensional rotor: the magnetic-like coupling
between the atoms of the BEC and the mechanical system results in the
interaction between a harmonic oscillator and one of the components of
the rotor. This allows to ``write'' information of the coherences
present in the cantilever state onto the state of the rotor, which can
then be read out using a technique based on the optical Faraday
effect. Our work provides a fully analytical framework for the
proposed protocol and discusses a number of relevant cases showing the
effectiveness of the scheme. The complexity of the problem, which
requires the management of a very large sector of the Hilbert space of
the cantilever-BEC system, demands the development of appropriate
methods to include the relevant sources of noise affecting the
device. A detailed treatment of this issue is left to future
work.


The remainder of this work is organized as follows. In Sec.~\ref{desc}
and \ref{meas} we introduce the setup, the magnetic-like interaction
Hamiltonian and, following Refs.~\cite{spinorham,mapping,rotor},
carefully guide through the formal mapping of the BEC onto a rotor. We
finally recast the BEC-cantilever coupling terms into the form of a
direct interaction between a harmonic oscillator and one of the
components of the BEC rotor.  In Sec.~\ref{dete} we propose a possible
detection scheme by means of the read out of such a rotor component
and apply our framework to a few relevant instances. In
Sec.~\ref{conc} we present our conclusions and briefly discusses a few
interesting open questions.


\section{The set up and the Hamiltonian}
\label{desc}

\begin{figure}[ht]
  \includegraphics[width=8 cm]{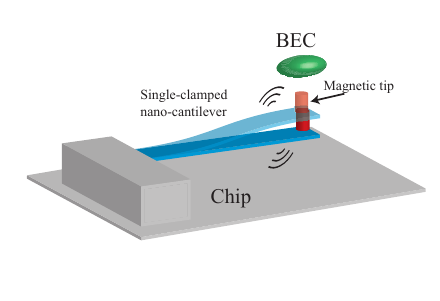}
  \caption{(Color online) Sketch of the set-up for BEC-based probing
    of mechanical coherences.  A BEC is placed in close proximity to a
    nano-mechanical cantilever endowed with a magnetic tip. The
    coupling between the magnetic field generated by the mechanical
    {\it quantum antenna} and the ultra-cold atoms embodies a
    mechanism for the effective probing of coherences in the state of
    the mechanical system.}
\label{fig:setup}
\end{figure}

We consider the setup sketched in Fig.~\ref{fig:setup}, which consists
of an on-chip single-clamped cantilever and a spinor BEC trapped in
close proximity to the chip and the cantilever. The latter is assumed
to be manufactured so as to accommodate at its free-standing end a
single-domain magnetic molecule (or {\it tip}). Technical details on
the fabrication methods of similar devices can be found in
Refs.~\cite{tip,hunger}, which have also been found to have very large
quality factors, which guarantee a good resolution of the rich variety
of modes in the cantilever's spectrum.
%
%
At room temperature, thermal fluctuations are able to (incoherently)
excite all flexural and torsional modes and in the following we assume
that a filtering process is put in place, restricting our observation
to a narrow frequency window, so as to select only a single mechanical
mode.


The second key element of our setup is a BEC of ${}^{87}$Rb atoms held
in an (tight) optical trap and prepared in the hyperfine level
$\ket{F=1}$.  As we assume the trapping to be optical, there is no
distinction between atoms with different quantum numbers $m_F=0,\pm1$
of the projections of the total spin along the quantization axis.
Moreover, for a moderate number of atoms in the condensate and a tight
trap, we can invoke the so-called single-mode approximation
(SMA)~\cite{SMAfail}, which amounts to considering the same spatial
distribution for all spin states.  These approximations will be made
rigorous and formal in the next Subsections.

\subsection{Hamiltonian of the system}

In the following we will briefly review the mapping of a spinor BEC
into a rotor~\cite{rotor}.  The Hamiltonian of a BEC in second
quantization reads~\cite{spinorham}
\begin{equation}
  \label{eq:ham}
  \begin{aligned}
    \hat{H}&=\!\!\sum_{\alpha} \int d{\bf x}\hat{\Psi}_{\alpha}^{\dagger}({\bf x})\hat{H}_{\alpha}^{0}\hat{\Psi}_{\alpha}({\bf x})\\
    &+\sum_{\alpha,\beta,\mu,\nu}\!\!G_{\alpha,\beta,\mu,\nu}\!\!\int d{\bf x}\hat{\Psi}_{\alpha}^{\dagger}({\bf x})\hat{\Psi}_{\beta}^{\dagger} ({\bf x})\hat{\Psi}_{\mu}({\bf x})\hat{\Psi}_{\nu}({\bf x})\\
  \end{aligned}
\end{equation}
where the second line of equation describes the particle-particle
scattering mechanism and $\hat{H}_{\alpha}^{0}{=}-(\hbar^2/2
m)\nabla^2+m (\omega^2(x^2+y^2)+\omega_z^2 z^2)/2$, $m$ is the mass of
the Rb atoms and the $\omega$ are the trapping frequencies in the
different spatial directions.  The subscripts $\alpha, \beta,\mu,\nu$
refer to different z-components of the single-atom spin states. Since
the scattering between two particles does neither change the total
spin nor its z-component, we can link the coefficients
$G_{\alpha,\beta,\mu,\nu}$ to the scattering lengths for the channels
with total angular momentum $F_{T}=0,2$. Thus, by making use of the
Clebsch-Gordan coefficients,
the full BEC Hamiltonian can be re-written as
\begin{equation}
  \label{eq:hamspin}
  \begin{aligned}
    &\hat{H}=\sum_{\alpha} \int d{\bf x} \;
    \hat{\Psi}_{\alpha}^{\dagger}({\bf x})\hat{H}_{\alpha}^{0}\hat{\Psi}_{\alpha}({\bf x})\\
    &+\frac{c_s}{2} \sum_{\alpha,\beta} \int d{\bf x} \; \hat{\Psi}_{\alpha}^{\dagger}({\bf x})\hat{\Psi}_{\beta}^{\dagger} ({\bf x})\hat{\Psi}_{\alpha}({\bf x})\hat{\Psi}_{\beta}({\bf x})\\
    &+\frac{c_a}{2}\!\!\sum_{\alpha,\beta,\alpha{'},\beta{'}}\!\!\int
    d{\bf x} \; \hat{\Psi}_{\alpha}^{\dagger}({\bf
      x})\hat{\Psi}_{\beta}^{\dagger} ({\bf x})({\bf
      F}_{\alpha,\beta}{\cdot}{\bf
      F}_{\alpha'\!,\beta})\hat{\Psi}_{\alpha{'}}({\bf
      x})\hat{\Psi}_{\beta{'}}({\bf x})
  \end{aligned}
\end{equation}
where $c_{s}{=}(g_0{+}2 g_2)/3$ and $c_{a}{=}(g_2{-}g_0)/3$ with
$g_{2j}{=}4 \pi \hbar^2 a_{2j}/m$ $(j{=}0,1)$ and $a_{2j}$ being the
scattering length for the $F_T=2j$
channel~\cite{scatteringlengths}. Here ${\bf F}$ is the vector of the
spin-$1$ matrices obeying the commutation relation $[{\bf F}^{i},{\bf
  F}^{j} ]=i \; \epsilon_{ijk} {\bf F}^{k}$ with $\epsilon_{ijk}$
being the Levi-Civita tensor.

As one can see from Eq.~\eqref{eq:hamspin}, if $c_a\approx 0$ ({\it
  i.e.}  if $g_0\approx g_2$) and/or the number of atoms is not too
large, the total Hamiltonian is symmetric in the three spin
components.  By assuming a strong enough optical confinement and a BEC
of a few thousand atoms, one can therefore think of the order
parameter as having a constant spatial distribution for all the three
species $m_F=0,\pm1$ and write
$\opx{\Psi}{x}{\alpha}{}{=}\funcx{\psi}{x}{}{}\hat{a}_{\alpha}{}$.
This is the so called single-mode approximation
(SMA)~\cite{SMAfail,spindynamics} which leaves the Hamiltonian in the
form
\begin{equation}
  \label{eq:single}
  \begin{aligned}
    \hat{H}&= \sum_{\alpha} \op{a}{\alpha}{\dagger}\op{a}{\alpha}{}
    +\frac{c{'}_s}{2}\sum_{\alpha,\beta}
    \op{a}{\alpha}{\dagger}\op{a}{\beta}{\dagger}
    \op{a}{\alpha}{}\op{a}{\beta}{}\\
    &+ \frac{c{'}_a}{2}\!\!\sum_{\alpha,\beta,\alpha{'},\beta{'}}\!\!
    ({\bf F}_{\alpha,\beta}\cdot{\bf F}_{\alpha{'},\beta{'}})\,
    \op{a}{\alpha}{\dagger}\op{a}{\alpha{'}}{\dagger}
    \op{a}{\beta}{}\op{a}{\beta{'}}{},
  \end{aligned}
\end{equation}
where we have defined $c'_{i}{=}c_{i} \int d{\bf x}\;
\modulo{\funcx{\psi}{x}{}{}}^{4}$.  As the distance $z_0$ between the
BEC and the magnetic tip can be in the range of a few $\mu$m (we take
$z_0=1.5 \mu$m in what follow) and the spatial dimensions of the BEC
are typically between tenths and hundredths of $\mu$m (we considered
$a_z=0.25 \mu$m and $a_r=0.09 \mu$m), the relative correction to the
magnetic field across the sample is of the order of
0.2, 
which is small enough to justify the SMA.  Moreover, in the
configuration assumed here, the system will be mounted on an atomic
chip, where the static magnetic field can be tuned by adding magnets
and/or flowing currents passing through side wires. Such a design can
compensate any distortions to the trapping potential induced by the
tip.

By introducing
$\hat{N}{=}\sum_{\alpha}\op{a}{\alpha}{\dagger}\op{a}{\alpha}{}$
%
%
and the angular momentum operators
$\op{L}{}{+}{=}\sqrt{2}(\op{a}{0}{\dagger}\op{a}{-1}{}+\op{a}{1}{\dagger}\op{a}{0}{})$
and
$\op{L}{z}{}{=}(\op{a}{1}{\dagger}\op{a}{1}{-}\op{a}{-1}{\dagger}\op{a}{-1}{})$~\cite{change},
we can rewrite Eq.~\eqref{eq:single} as
$\hat{H}{=}\op{H}{A}{}+\op{H}{S}{}$, where we have explicitly
identified a symmetric part $\op{H}{S}{}=\mu
\hat{N}-c'_{s}\hat{N}(\hat{N}-1)$ and an antisymmetric one
$\op{H}{A}{}=c'_{a}(\hat{L}^2-2\hat{N})$.
It is important to remember that such a mapping is possible due to the
assumption of a common spatial wave function for the three spin
components. As long as the antisymmetric term is small enough, this is
not a strict constraint.  By exploiting Feshbach resonances~\cite{feshbach}, it is
possible to adjust the couplings $g_0$ {and} $g_2$ in such a way that
$g_0{\approx}g_2$, which allows for the possibility to increase the
number of atoms in the BEC, still remaining within the validity of the SMA.

We now consider the BEC interaction Hamiltonian when an external
magnetic field is present.  Due to its magnetic tip, the cantilever
produces a magnetic field and we assume that only one mechanical mode
is excited, so that the cantilever can be modeled as a single quantum
harmonic oscillator
whose annihilation (creation) operator we call $\op{b}{c}{}$
($\op{b}{c}{\dagger}$). By allowing the tip to have an intrinsic
magnetization, we can split the magnetic field into a static
contribution ${\bf B}^{0}$ and an oscillating one $\delta \op{{\bf
    B}}{}{}$ that arises from the oscillatory behavior of the
mechanical mode.  The physical mechanism of interaction is
Zeeman-like, {\it i.e.} each atom experiences a torque which tends to
align its total magnetic moment to the external magnetic field. The
Hamiltonian for a single atom can be written as
\begin{equation}
 \label{eq:zee}
 \op{H}{Z}{(1)}=-{\boldsymbol \mu}{\cdot}{\bf B}= ({g\mu_{B}}/{\hbar})\op{{\bf S}}{}{(1)}{\cdot}{\bf B},
\end{equation}
where $\mu_{B}$ is the Bohr magneton, $\op{{\bf S}}{}{(1)}$ is the
spin operator vector for a single atom and $g$ is the gyromagnetic
ratio. In line with Ref.~\cite{gyro}, we adopt the convention that $g$
and ${\bf \mu}$ have opposite signs.  The total interaction
Hamiltonian is then given by the sum over all the atoms.
By taking the direction of ${\bf B}^{0}$ as the quantization axis
(z-axis) and the x-axis in the direction of $\mean{\delta \op{{\bf
      B}}{}{}}{}$, the magnetic Zeeman-like Hamiltonian is
\begin{equation}
 \label{eq:zeeang}
 \op{H}{Z}{}=g\mu_{B}B_{z}^{0}\op{L}{z}{}+g\mu_{B}G_ca_{c}(\op{b}{c}{\dagger}+\op{b}{c}{})\op{L}{x}{},
\end{equation}
where we have used $\delta \op{{\bf B}}{}{}= G_c
a_{c}(\op{b}{c}{\dagger}+\op{b}{c}{})\overline{{\bf x}}$ with $G_c=3\mu_0|{\boldsymbol \mu}_c|/(4\pi z_0^4)$
being the gradient of the magnetic field produced by the tip at a distance
$z_0$,
$\overline{{\bf x}}$ the unit vector along the x-axis,
$a_{c}=\sqrt{\hbar/(2 m_{e} \omega_{c})}$
and $m_e$ the effective mass of the cantilever~\cite{mass}.
The full Hamiltonian of the BEC-cantilever system is thus $\op{H}{}{}=
\op{H}{BEC}{0}+\op{H}{c}{0}+\op{H}{I}{}$ with
\begin{equation}
 \begin{aligned}
 \label{eq:totham}
  &\op{H}{BEC}{0}=\mu \hat{N}{-}c'_{s}\hat{N}(\hat{N}-1){+}
 c'_{a}(\hat{L}^2-2\hat{N}){+}g\mu_{B}B_{z}^{0}\op{L}{z}{},\\
 &\op{H}{c}{0}=\hbar\omega_{c}\op{b}{c}{\dagger}\op{b}{c}{},\\
 &\op{H}{I}{}=g\mu_{B}G_c a_{c}(\op{b}{c}{\dagger}+\op{b}{c}{})\op{L}{x}{}.
 \end{aligned}
\end{equation}
It has been shown in Refs.~\cite{spinorham,spindynamics} that
$\op{H}{BEC}{0}$ with $B_{z}^{0}=0$ allows for an interesting dynamics
of the populations of the three spin states, which undergo Rabi-like
oscillations, thus witnessing the coherence properties of the BEC.


\subsection{Mapping into a rotor}

While the Hamiltonian above is rather appealing, it is not yet in a
form that is of use for our application. In fact, let us consider the
natural basis to describe the system, {\it i.e.}~the one spanned by
$\ket{L,L_z}$, which are the common eigenstates of $\hat L$ and $\hat
L_z$.  Due to the coherence in the state of the BEC, we cannot fix the
quantum number $L$, since, for instance, if the BEC is in an
eigenstate of $\op{L}{z}{}$ with $L_z=0$, then the state has the form
$\sum_{L=0}^{N}c_{L} \ket{L,0}$.  Tracking the evolution induced by
Eq.~(\ref{eq:totham}) on such a superposition is a non trivial problem
since for $N\gg1$ the accessible region of the Hilbert space becomes
quite large. Nevertheless, the problem can be tackled by
the formal mapping of the BEC into a quantum rotor.  In the following,
we briefly discuss the basic ideas of this mapping as given in
Ref.~\cite{rotor}.  Since we work with a fixed number of particles,
the state of the BEC can be decomposed as
\begin{equation} 
  \label{gen}
  \sum_{n'_{0,\pm1}}C_{n'_{0,\pm1}}(\op{a}{1}{\dagger})^{n'_{1}}(\op{a}{0}
  {\dagger})^{n'_{0}}(\op{a}{-1}{\dagger})^{n'_{-1}}\ket{0}
\end{equation}
where the sum is performed over all sets of labels $\{n'_{0,\pm1}\}$
such that $n'_{0}{+}n'_{-1}{+}n'_{1}{=}N$.  Let us now introduce the
Schwinger-like operators
$\op{b}{x}{}{=}(\op{a}{-1}{}{-}\op{a}{1}{})/\sqrt{2}$,
$\op{b}{y}{}{=}(\op{a}{1}{}{+}\op{a}{-1}{})/(i \sqrt{2})$,
$\op{b}{z}{}{=}\op{a}{0}{}$ such that
$[\op{b}{\alpha}{},\op{b}{\beta}{}]{=}0$,
$[\op{b}{\alpha}{},\op{b}{\beta}{\dagger}]{=}\delta_{\alpha,\beta}$~\cite{rotor}.
The generic BEC state in Eq.~\eqref{gen} can now be written as
$\ket{\Omega_{N}}{=}\frac{1}{\sqrt{N!}}({\bf \Omega}{\cdot}{\bf
  \op{b}{}{\dagger}})^{N}\ket{0}$ with ${\bf
  \Omega}{=}(\cos\phi\sin\theta,\sin\phi\sin\theta,\cos\theta)$.  By
varying $(\theta,\phi)$ and thus the position vector $\vert{\bf
  \Omega}\rangle$ on the unit sphere, it is possible to recover any
superposition for the state of a single atom among the states with
$m_z=0,\pm 1$.  Any state with a fixed number of particles in the
bosonic Hilbert space can then be written as $\ket{\Psi}=\int
d\Omega\; \ket{{\bf \Omega}_N} \psi({\bf \Omega})$ where $\psi({\bf
  \Omega})$ is the wave function of the rotor we are looking for to
complete the mapping.  The next step is then to find the form of the
Hamiltonian in this space.  According to Ref.~\cite{rotor}, a
sufficient criterion for the two dynamics to be equivalent is the
existence of a Hamiltonian operator $\hat{\cal H}$ in the Hilbert
space of the rotor such that $\op{H}{}{}\ket{\Psi}=\int
d\Omega\ket{{\bf \Omega}_N} \hat{\cal H}\psi(\bf\Omega)$. The explicit
form of $\hat{\cal H}$ can in fact be found by a straightforward
calculation that leads to the expressions of the $z$ and $x$
components of the angular momentum operator of the form
\begin{equation}
  \begin{aligned}
    \op{L}{z}{}&=-i
    (\op{b}{x}{\dagger}\op{b}{y}{}-\op{b}{y}{\dagger}\op{b}{x}{})=
    -i \overline{{\bf z}}\cdot({\bf \Omega}\times\nabla)=\frac{1}{\hbar}\overline{{\bf z}} \cdot \op{\mathcal{L}}{}{}=-i \partial_{\phi},\\
    \op{L}{x}{}&=\frac{1}{2}(\op{b}{z}{\dagger}\op{b}{x}{}-\op{b}{x}{\dagger}\op{b}{z}{})+\frac{i}{2} \;(\op{b}{z}{\dagger}\op{b}{y}{}-\op{b}{y}{\dagger}\op{b}{z}{})=-i\overline{{\bf x}}\cdot({\bf \Omega}\times\nabla)\\
    &=\frac{1}{\hbar}\overline{{\bf x}} \cdot
    \op{\mathcal{L}}{}{}=i(\sin\phi\;\partial_{\theta}{+}\cot\theta\cos\phi\partial_{\phi}).
  \end{aligned}
\end{equation}
After discarding an inessential constant term, the Hamiltonian that we
are looking for reads
$\op{\mathcal{H}}{}{}=\op{\mathcal{H}}{R}{0}+\op{\mathcal{H}}{c}{0}+\op{\mathcal{H}}{I}{}$
with
\begin{equation}
 \begin{aligned}
 \label{eq:rotorham}
	\op{\mathcal{H}}{R}{0}&= \;c'_{a}\op{\mathcal{L}}{}{2}+(g{\mu_{B}}/{\hbar})B_{z}^{0}\;\op{\mathcal{L}}{z}{},\\
	\op{\mathcal{H}}{c}{0}&={\op{p}{c}{2}}/{2m_{e}}+m_{e}\omega_{c}^{2}\op{q}{c}{2}/2,\\
	\op{\mathcal{H}}{I}{}&=({g\mu_{B}}/{\hbar})G_c\op{q}{c}{}\op{\mathcal{L}}{x}{}.
 \end{aligned}
\end{equation}
In Eq.~\eqref{eq:rotorham} we have introduced, for convenience, the
cantilever's position and momentum operators $\hat{q}_c=\sqrt{\hbar/(2
  m\omega_c)}(\hat{b}_c+\hat{b}^\dag_c)$ and
$\hat{p}_c=i\sqrt{\hbar{m}\omega_c/2}(\hat{b}^\dag_c-\hat{b}_c)$. We
are now in a position to look at BEC-cantilever joint dynamics.  In
particular we will focus on the detection of the cantilever properties
by looking at the BEC spin dynamics.

\section{Probing quantum coherences}
\label{meas}

\subsection{Dynamics}

The form of the interaction Hamiltonian $\op{\mathcal{H}}{I}{}$ allows
for the measurement of any observable whose corresponding operator on
the Hilbert space can be expressed as a function of $\op{q}{c}{}$ and
$\op{p}{c}{}$ with no back action on the cantilever dynamics.
Moreover, when there is no magnetic field, the ground state of a
``ferromagnetic'' ({\it i.e.}~$c_2<0$) spinor BEC is such that all the
atomic spins are aligned along a direction resulting from a
spontaneous symmetry breaking process~\cite{spinorham}. Under the
effects of the cantilever antenna, two preferred directions are
introduced in the system: the $z$-direction along which we have the
static magnetic field and the $x$-direction defined by the oscillatory
component.  The interplay between these two competing magnetic fields
is responsible for a ``gyroscopic'' motion of the rotor about the
$z$-axis, exactly as in a classical spinning top. By looking at the
way the rotor undergoes such a gyromagnetic motion, we can gather
information about the properties of the cantilever state. We notice
that a similar approach has been used to show the resonant coupling of
an atomic sample of ${}^{87}$Rb atoms with a magnetic tip similar to
the one considered here~\cite{magntip}. 
 
In order to understand the mechanism, let us look at the time
evolution of the operator $\opt{\mathcal{L}}{t}{x}{}$.  We take an
initial state of the form
\begin{equation}
  \label{eq:initial}
  \ket{\Psi(0)}=\sum_{n}C_{n}\ket{E_{n}}\int_{\Sigma_1}d\Omega\;\psi(\Omega)\ket{\Omega}
\end{equation}
where $\Sigma_{1}$ is the unit sphere and $\ket{E_{n}}$ are the energy
eigenvalues for the harmonic oscillator such that
$\op{\mathcal{H}}{c}{0}\ket{E_{n}}{=}E_{n}\ket{E_{n}}$. In the
Heisenberg picture, the mean value of the x-component of the angular
momentum is
\begin{widetext}
 \begin{equation}
   \begin{aligned}
     \label{eq:mean}
     \mean{\opt{\mathcal{L}}{t}{x}{}}{}&=\braketM{\Psi(0)}{e^{i  \frac{\op{\mathcal{H}}{}{}}{\hbar}t}\opt{\mathcal{L}}{0}{x}{}e^{-i  \frac{\op{\mathcal{H}}{}{}}{\hbar}t}}{\Psi(0)}\\
     &=\int_{\Sigma_1,q,q'}d\Omega dq dq' \sum_{n,m}C_{m}^{*}C_{n}e^{-i  \omega_{n,m} t}\phi_{m}^{*}(q')\phi_{n}(q) \braket{q'}{q} \psi^{*}(\Omega)\left({e^{i  \frac{\op{\mathcal{H}_I}{}{}+\op{\mathcal{H}}{R}{0}}{\hbar}t}\opt{\mathcal{L}}{0}{x}{}e^{-i  \frac{\op{\mathcal{H}_I}{}{}+\op{\mathcal{H}}{R}{0}}{\hbar}t}}\right)\psi(\Omega)\\
     &=\int_{q}dq \sum_{n,m}C_{m}^{*}C_{n}e^{-i \omega_{n,m}
       t}\phi_{m}^{*}(q)\phi_{n}(q) \int_{\Sigma_1}d\Omega
     \;\psi^{*}(\Omega)\bigg({e^{i
         \frac{\op{\mathcal{H}_I}{}{}+\op{\mathcal{H}}{R}{0}}{\hbar}t}\opt{\mathcal{L}}{0}{x}{}e^{-i
         \frac{\op{\mathcal{H}_I}{}{}+\op{\mathcal{H}}{R}{0}}{\hbar}t}}\bigg)[\psi(\Omega)]
   \end{aligned}
 \end{equation}
\end{widetext}
where we have used the closure relation
$\int_{q}\ketbra{q}{q}=\openone$ twice
and introduced $\phi_{n}(q){=}\braket{q}{E_{n}}$ and
$\omega_{n,m}{=}\omega_{c}(n-m)$.  By setting
$\Omega_q{=}\sqrt{(g\mu_B/\hbar)^2[(B_z^0)^2+G_c^2 q^2]}$, the
time-evolved x-component of the angular momentum operator is
\begin{equation}
  \label{eq:lx}
\begin{aligned}
  &\opt{\mathcal{L}}{t}{x}{}
  =\frac{g^2 \mu_B^2}{\hbar^2\Omega^{2}(q)}\left[(B_{z}^{0})^{2}\cos(\Omega_qt)+G_{c}^{2}q^{2}\right]\opt{\mathcal{L}}{0}{x}{}\\
  &{+}\frac{g \mu_B B_z^0 }{\hbar\Omega_q}\sin(\Omega_qt)\opt{\mathcal{L}}{0}{y}{}{+}\frac{g^2 \mu_B^2 B_z^0 G_c q}{\hbar^2\Omega^{2}(q)}\left[1{-}\cos(\Omega_qt)\right]\opt{\mathcal{L}}{0}{z}{}\\
  &=a_1(q,t)\opt{\mathcal{L}}{0}{x}{}+a_2(q,t)\opt{\mathcal{L}}{0}{y}{}+a_3(q,t)\opt{\mathcal{L}}{0}{z}{}.
\end{aligned}
\end{equation}
Comparing Eqs.~\eqref{eq:mean} and \eqref{eq:lx} we find
$\langle{\opt{\mathcal{L}}{t}{x}{}}\rangle{=}\sum_{j=x,y,z}A_j(t)L_j^0
$,
where
\begin{equation}
  \begin{aligned}
    L_j^{0}&=\int_{\Sigma_1}d\Omega \;\psi^{*}(\Omega)\opt{\mathcal{L}}{0}{j}{}[\psi(\Omega)],\\
    A_j(t)&=\sum_{n,m}C_{m}^{*}C_{n}e^{-i \omega_{n,m} t}\int_{q}dq
    \phi_{m}^{*}(q)\phi_{n}(q) a_j(q,t).
\end{aligned}
\end{equation}
If the cantilever is initially prepared in the general mixed state $\rho_c(0)=\sum_{n}C_{n,m}\ketbra{E_n}{E_m}$,
a similar expression for the mean value of $\opt{\mathcal{L}}{t}{x}{}$ is found, where now
$A_j=\sum_{n,m}e^{-i  \omega_{n,m} t}C_{n,m}\int_{q}dq \phi_{m}^{*}(q)\phi_{n}(q) a_j(q,t)$.

\begin{figure}[t]
  \includegraphics[width=\linewidth]{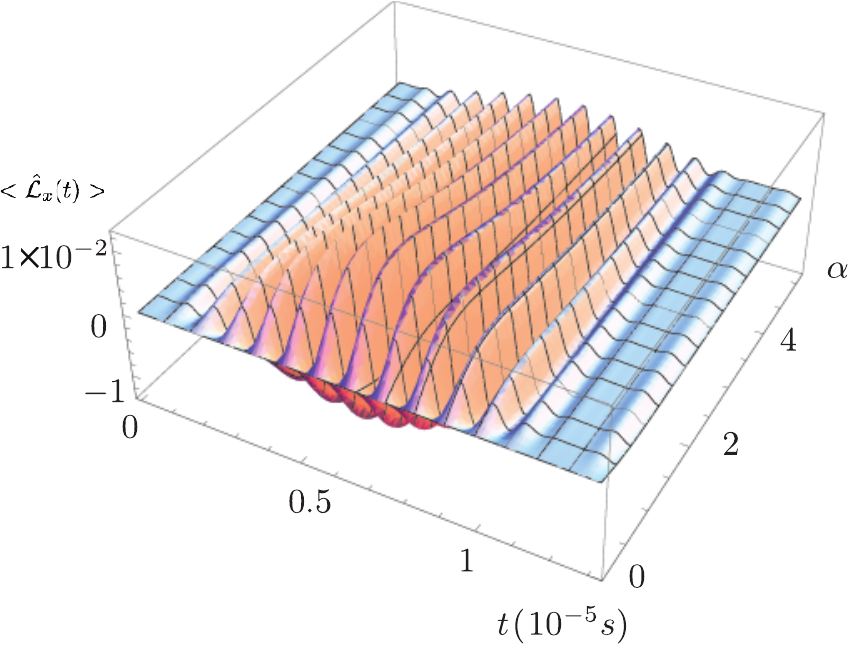}
  \caption{(Color online) Mean value of $\opt{\mathcal{L}}{t}{x}{}$
    for a cantilever in the initial state as given by
    Eq.~(\ref{eq:initial}) with $C_0{=}C_1/\alpha{=}1/\sqrt{1+\alpha^2}$
    and $C_n=0$ otherwise.  The BEC consists of $N{=}10^3$
    ${}^{87}\text{Rb}$ atoms and
    $\langle{\opt{\mathcal{L}}{0}{x,y}{}}\rangle=0, \;
    \langle{\opt{\mathcal{L}}{0}{z}{}}\rangle=100$. 
    {We have used $B_z^0{=}3\times 10^{-6} \mu T$ 
            and $G_c\approx 1.8 \times 10^3 \mu T/\mu$m}.}
\label{fig:coh01}
\end{figure}
As the qualitative conclusions of our analysis do not depend upon the
initial value of the angular momentum component of the spinor, in what
follows we shall concentrate on an illustrative example that allows us
to clearly display our results. We thus consider, without affecting
the generality of our discussions,
$\mean{\opt{\mathcal{L}}{0}{x,y}{}}{}
=0$ {and} $\mean{\opt{\mathcal{L}}{0}{z}{}}{}=100$.  When the
cantilever and the BEC are uncoupled, we should expect
$\mean{\opt{\mathcal{L}}{t}{x}{}}{}$ to oscillate at the Larmor
frequency $\omega_L=g \mu_B B_z^0$ and with an amplitude independent
of $\mean{\opt{\mathcal{L}}{0}{x}{}}{}$.  The BEC-cantilever coupling
introduces a modulation of such oscillations and in the following we
will demonstrate that the analysis of such oscillatory behavior is
indeed useful to extract information on the state of the cantilever.

\begin{figure}[t]
  \includegraphics[width=\linewidth]{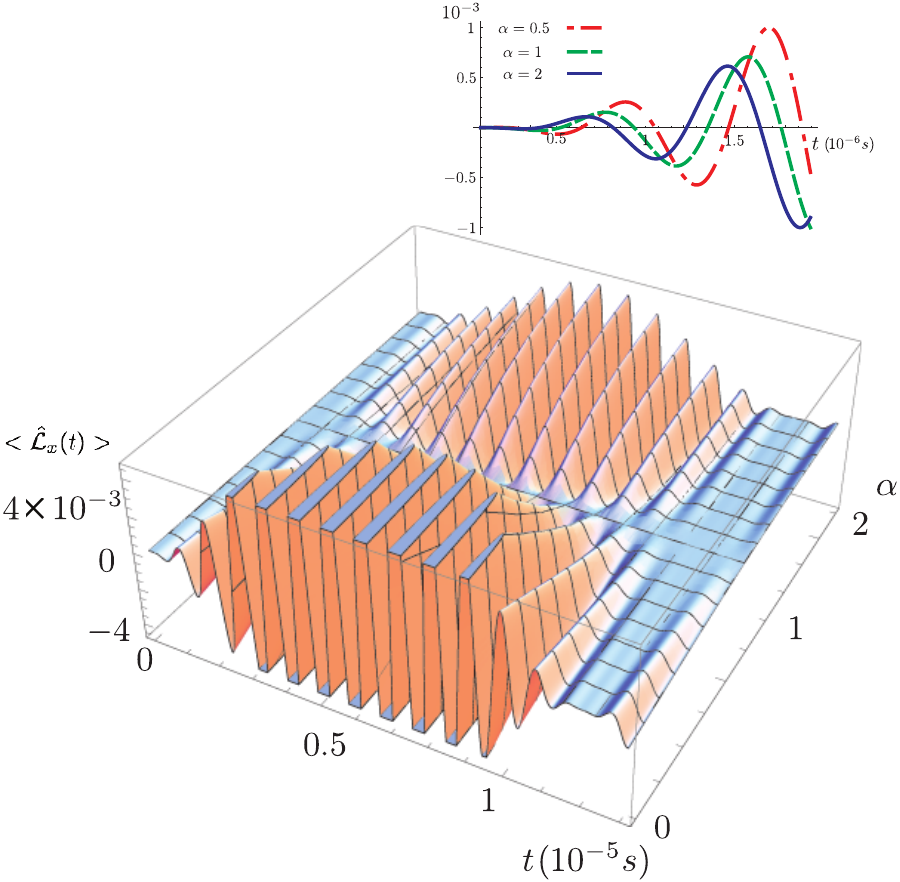}
  \caption{(Color online) Mean value of $\opt{\mathcal{L}}{t}{x}{}$
    for a cantilever in the initial state as given by
    Eq.~(\ref{eq:initial}) with $C_0=C_1=C_2/\alpha=1/\sqrt{2+\alpha^2}$ 
    and $C_n=0$ otherwise. The BEC parameters are the same as in
    Fig.~\ref{fig:coh01}. The inset shows that the change in $|\alpha|$ amounts to a shift of the oscillations [we have taken $=e^{i \pi/6}(0.5, 1, 2)$].}
\label{fig:coh012}
\end{figure}

\begin{figure}[h]
\includegraphics[width=8 cm]{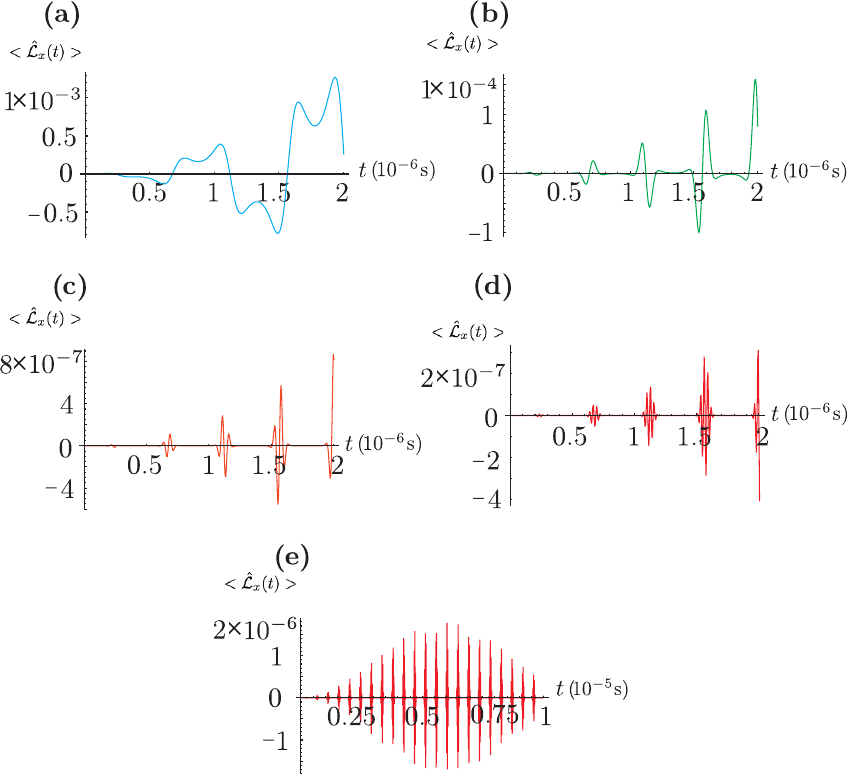}
\caption{(Color online) Time evolution of $\op{\mathcal{L}}{x}{}$ for
  a coherent initial state of the cantilever with $|\alpha|^2=1\;
  \;\text{{\bf (a)}},5\;\text{{\bf (b)}},15\;\text{{\bf (c)}},20\;
  \text{{\bf (d)}}$.  For the same parameter as in $\text{{\bf (d)}}$,
  the plot $\text{{\bf (e)}}$ shows that the carrier frequency
  $\omega_L$ is not significantly affected.}
\label{fig:coherent}
\end{figure}

We first consider the case of a cantilever initially prepared in a
superposition of a few eigenstates of the free Hamiltonian
$\op{\mathcal{H}}{c}{0}$, as in Eq.~(\ref{eq:initial}).  In
Fig.~\ref{fig:coh01} we show the mean value of
$\opt{\mathcal{L}}{t}{x}{}$ as a function of the coherence between the
states with quantum number $n=0$ and $n=1$, {\it i.e.}~a state having
$C_0=C_1/\alpha=1/\sqrt{1+\alpha^2}$ and $C_n=0$ otherwise.  One can
see a clear modulation of the behavior of $\langle\hat{\cal
  L}_x(t)\rangle$: a close inspection reveals that the carrier
frequency $\omega_L$ is modulated by the frequency $\omega_{0,1}$. In
reality, the Larmor frequency is renormalized as can be seen by the
expression for $\Omega_q$. However, as we have taken $G_c a_c\ll B_z^0$,
one can safely assume that the carrier frequency is very close to $
\omega_L$. Moreover, the maximum of the function is found at
$C_{0,1}=1/\sqrt{2}$, which maximizes the coherence between the two
states and thus the effect of the modulation.  For symmetry reasons,
the modulation described is not visible if the cantilever is prepared
in a superposition of phonon eigenstates whose quantum numbers are all
of the the same parity (such as a single-mode squeezed state).  In
this case, in fact, the function entering the integral over $q$ in
$A_3$ is antisymmetric, thus making it vanish.  In
Fig.~\ref{fig:coh012}, $\langle\opt{\mathcal{L}}{t}{x}{}\rangle$ is
shown for an initial state of the cantilever having
$C_{0,1}=C_2/\alpha=1/\sqrt{2+\alpha^2}$ and $C_n=0$ otherwise.  It is
worth noticing that one can identify two regions of oscillations
separated by the line of nodes at $\alpha=1$ where $C_0=C_1=C_2$. We
can understand this behavior by studying the amplitudes of oscillation
in three $\alpha$-dependent regions. For $\alpha<1$, the main
modulation frequency is given by $\omega_{0,1}$ and the role of the
third state is to modify the amplitude of the oscillations [see
Fig.~(\ref{fig:coh012})].  At $\alpha=1$ a destructive interference
takes place and the amplitude drops down.  For $\alpha>1$ the
frequency $\omega_{1,2}$ enters into the evolution of
$\langle\opt{\mathcal{L}}{t}{x}{}\rangle$ (for parity reasons, the
term with frequency $\omega_{0,2}$ has no role) and determines a phase
shift of the oscillation
fringes.
It is interesting to observe that if the initial state of the
cantilever is purely thermal, $\mean{\opt{\mathcal{L}}{t}{x}{}}{}$
does not oscillate: only quantum coherence in the state of the
mechanical system give rise to oscillatory behaviors and their
presence is well signaled by the pattern followed by the angular
momentum of the spinor-BEC.
Although the examples considered so far have been instrumental in
explaining the connections between the properties of the cantilever
and the dynamics of the spinor's degrees of freedom, they are
unfortunately currently far from being realistic. We will therefore
now consider closer-to-reality example of a pure state that is likely
to be achieved soon. Given the impressive advances in the control and
state-engineering of micro and nano-mechanical systems, we will
consider the cantilever to be prepared in a coherent state with an
average phonon number $n_{ph}$~\cite{commentocohe} .
In Fig.~\ref{fig:coherent} we show the time evolution of
$\opt{\mathcal{L}}{t}{x}{}$ for $|\alpha|^2=1$ [panel \text{{\bf
    (a)}}], $5$ [panel \text{{\bf (b)}}], $15$ [panel \text{{\bf (c)}}],
and $20$ [panel \text{{\bf (d)}}]. One can see that, depending on the
mean number of phonons initially present in the mechanical state, new
frequencies are introduced in the dynamics of the device: the larger
$|\alpha|^2$, the larger the number of frequencies involved due to the
Poissonian nature of the occupation probability distribution of a
coherent state.
In Fig.~\ref{fig:coherent} {\bf (e)}, which addresses the case of
$|\alpha|^2=20$, the study of the dynamics at long evolution times
reveals that the carrier frequency is unaffected, for all practical
purposes, while the large number of frequencies entering in the
evolution gives rise to series of beats occurring at different time
scales.

\subsection{Detection scheme}
\label{dete}

To read out the information imprinted on the rotor, one can make use
of the Faraday-rotation effect, which allows to measure one component
of the the angular momentum of the BEC with only a negligible back
action on the condensate itself. It is well-known from classical
optics that the linear polarization of an electromagnetic field
propagating across an active medium rotates with respect to the
direction it had when entering the medium itself. This is the essence
of the Faraday-rotation effect, which can be understood by decomposing
the initial polarization in terms of two opposite circularly-polarized
components experiencing different refractive
indices~\cite{polartensor}: by going through the medium, the two
components acquire different phases, thus tilting the resulting
polarization.

In the case of an ultra-cold gas, an analogous rotation of the
polarization of a laser field propagating across the BEC is due to the
interaction of light with the atomic spins. If the spins are randomly
oriented the net effect is null, while for spins organized in
clusters, the effect can indeed be measured. It has been shown in
Refs.~\cite{faradayQNM, faradayback} that the back-action on the BEC
induced by this sort of measurements is rather negligible. In recent
experiments non destructive measurements on a single BEC of
${}^{23}$Na atoms have been used to show the dynamical transition
between two different regions of the stability diagram of the
system~\cite{faradayNa}. This method can thus be effectively used to
determine the dynamics of the angular momentum components of the rotor
BEC and thus indirectly witness the presence of coherences in the
state of the cantilever.  {Moreover, as shown in
  Ref.~\cite{faradayback}, the signal to noise ratio (SNR) is
  proportional to $\sqrt{\tau_{pd}/\tau_s}$ where $\tau_{pd}$ is the
  characteristic time for the response of the photo-detector and
  $\tau_s$ is the average time between consecutive photon-scattering
  events. In order to be able to detect two distinct events on a time
  scale $\tau$ we thus need $\tau_{pd}<\tau<\tau_s$ to hold. This
  condition states that the number of scattered photons has to be
  small enough during the time $\tau$ over which the dynamics we want
  to resolve occurs. On the other hand the detector ``death time''
  should be smaller than the typical evolution time.  While $\tau_s$
  can be easily tuned by adjusting the experimental working point,
  ultrafast photo-detectors of the latest generation have response
  time $\tau_{pd}$ of a few $ps$. As in our scheme we have
  $\tau\in[10^{-8},10^{-5}]$s, the proposed coherence-probing method
  appears to be within reach.}


\section{Conclusion}
\label{conc}

We have considered a mechanical cantilever equipped with a magnetic
tip interacting with a spinor Bose-Einstein condensate (BEC) held in
an optical trap.  The tip produces a magnetic field made up of two
components, namely a static one along the tip's natural anisotropic
axes and one perpendicular to it due to the cantilever's oscillations.
By exploiting the mapping of a spinor BEC into a rotor
model~\cite{rotor} it is possible to take into account its quantum
properties which would have been missed in a mean field theory
approach.  The BEC is thus mapped onto a quantum gyroscope undergoing
a precession about the direction of the magnetic tip's static field.
We have assumed that the cantilever has been cooled down~\cite{cool,cool2} to a quantum
regime and described it as a quantum harmonic oscillator.  We have
shown that it is possible to detect the presence of quantumness in the
cantilever state in the form of superposition of different eigenstate
of the harmonic oscillator.  The way to do this is to look at the
gyroscopic precession by using Faraday spectroscopy, which in turn
only minimally disturbes the BEC dynamics, thus allowing for a
continuous probing of the system.  Even though we have restricted our
analysis to a cantilever equipped with a magnetic molecule it is posible 
to generalize this scheme to other sorts of mesoscopic magnetic system such as nanotubes.

\noindent
{\it Note:} {During completion of this work, a related investigation reporting on the measurement back-action 
on a vibrating membrane coupled to a BEC has appeared~\cite{meystre}. While the detailed context and general approach 
 differ from ours, this work reinforces the idea that quantum coherence in mechanical systems can be reliably probed 
 by ultracold atomic systems.}

\acknowledgments This work was supported by IRCSET through the Embark
Initiative RS/2009/1082, Science Foundation Ireland under grant
numbers 05/IN/I852 and 10/IN.1/I2979, EUROTECH, and the UK EPSRC
(EP/G004579/1).


\begin{thebibliography}{}
\bibitem{change}We have inverted the direction of the raising operator
  as well as the sign of $\op{L}{}{z}$ with respect to
  Ref.~\cite{spinorham}.  This does not change the expression for the
  Hamiltonian in terms of angular momentum operators.
\bibitem{mass} The effective mass $m_{e}$ represents the mass involved
in the oscillation of the mode considered, which might be different from the total 
mass of the cantilever.
\bibitem{commentocohe} A mechanical coherent state will be general by displacing, with an intense laser field, the ground state of a cantilever. This is a realistic expectation: current state of the art experiments are only a few quanta away from such an achievement~\cite{cool2}.
\end{thebibliography}
\end{document}